\documentclass[12pt]{iopart}

\usepackage{iopams}
\usepackage{setstack}
\usepackage{graphicx}
\usepackage{hyperref}
\hypersetup{
  colorlinks,%
  citecolor=black,%
  filecolor=black,%
  linkcolor=black,%
  urlcolor=blue
  }

\usepackage[square,sort,compress,comma,numbers]{natbib}
  
\usepackage{float}
\usepackage{multirow}
\usepackage{xspace}

\usepackage{bm}

\newcommand*{\sqs}{\ensuremath{\sqrt{s}}\xspace}
\newcommand*{\sqsn}{\ensuremath{\sqrt{s_{\rm NN}}}\xspace}

\newcommand*{\Nch}{\ensuremath{N_\mathrm{ch}}\xspace}

\newcommand*{\pT}{\ensuremath{p_\mathrm{T}}\xspace}
\newcommand*{\mT}{\ensuremath{m_\mathrm{T}}\xspace}

\newcommand{\dd}{\ensuremath{\mathrm d}\xspace}

\usepackage{orcidlink}

\newcommand{\orcidA}{\orcidlink{0000-0002-2420-7650}} 
\newcommand{\orcidB}{\orcidlink{0000-0003-2849-0120}} 
\newcommand{\orcidC}{\orcidlink{0000-0001-9223-6480}} 
\newcommand{\orcidD}{\orcidlink{0000-0003-3706-5265}} 
\begin{document}
\title{How far can we see back in time in high-energy collisions using charm hadrons?}

\author{L\'aszl\'o Gyulai\orcidA{}$^{1, 2}$, G\'abor B\'ir\'o\orcidB{}$^{2, 3}$, R\'obert V\'ertesi\orcidD{}$^{2,\dag}$, Gergely G\'abor Barnaf\"oldi\orcidC{}$^2$}
\address{$^1$ Budapest University of Technology and Economics, Műegyetem rkp. 3., H-1111 Budapest, Hungary.}
\address{$^2$ HUN-REN Wigner Research Center for Physics, 29--33 Konkoly--Thege Mikl\'os Str., H-1121 Budapest, Hungary.}
\address{$^3$ ELTE E\"otv\"os Lor\'and University, Institute of Physics, 1/A P\'azm\'any P\'eter S\'et\'any, H-1117  Budapest, Hungary.}
\address{$^\dag$ corresponding author}
\eads{\mailto{gyulai.laszlo@wigner.hun-ren.hu}, \mailto{biro.gabor@wigner.hun-ren.hu}, \mailto{vertesi.robert@wigner.hun-ren.hu}, \mailto{barnafoldi.gergely@wigner.hun-ren.hu}}

\vspace{10pt}
\begin{indented}
\item[]\today
\end{indented}

\begin{abstract}
We use open charm production to estimate how far we can see back in time in high-energy hadron-hadron collisions. We analyze the transverse momentum distributions of the identified D mesons from pp, p--Pb and A--A collisions at the ALICE and STAR experiments covering the energy range from $\sqsn = 200$~GeV up to 7~TeV. Within a non-extensive statistical framework, the common Tsallis parameters for D mesons represent higher temperature and more degrees of freedom than that of light-flavour hadrons. Assuming Bjorken-expansion, the production of D mesons corresponds to a significantly earlier proper time, $\tau_{\rm D} = (0.18 \pm 0.06) \tau_{\rm LF}$.
\end{abstract}

\vspace{2pc}

\submitto{\JPG}

\section{Introduction}
\label{sec:nonext}

In the theory of strong interaction, quantum chromodynamics (QCD), quarks and gluons at low energy exist in colour singlet states, confined into hadrons. However, under the extreme conditions of high-energy heavy-ion collisions, quark--gluon plasma (QGP) can be formed. This state of matter features deconfined quarks and gluons with colour degrees of freedom. It is believed that such a plasma existed microseconds after the Big Bang, providing a glimpse into the primordial Universe. Over recent decades, experiments conducted at the Relativistic Heavy Ion Collider (RHIC) and the Large Hadron Collider (LHC) have been pivotal in advancing our understanding of the QGP by systematically measuring many of its fundamental properties~\cite{Evans:2008zzb, PHENIX:2004vcz, STAR:2005gfr, ALICE:2022wpn, Busza:2018rrf, Dokshitzer:2001zm}. These experiments have observed a strong collectivity in high-energy heavy-ion collisions, attributed to a strongly coupled QGP. Recently, collective phenomena have also been observed in smaller collision systems, such as proton-nucleus (p–A) or even proton-proton (pp) collisions~\cite{CMS:2010ifv,ATLAS:2012cix, ALICE:2012eyl}. Similarly, an enhancement in strangeness production has been seen in pp systems~\cite{ALICE:2016fzo}. Whether QGP droplets can form in small systems, or if the observed collectivity is a result of complex vacuum-QCD effects, remains an open question. Describing these observations in a unified framework poses a challenge for theories~\cite{PHENIX:2018lia, Ortiz:2016kpz, Nagle:2018nvi}.

\begin{figure}[t]
  \centering
  \includegraphics[width=\linewidth]{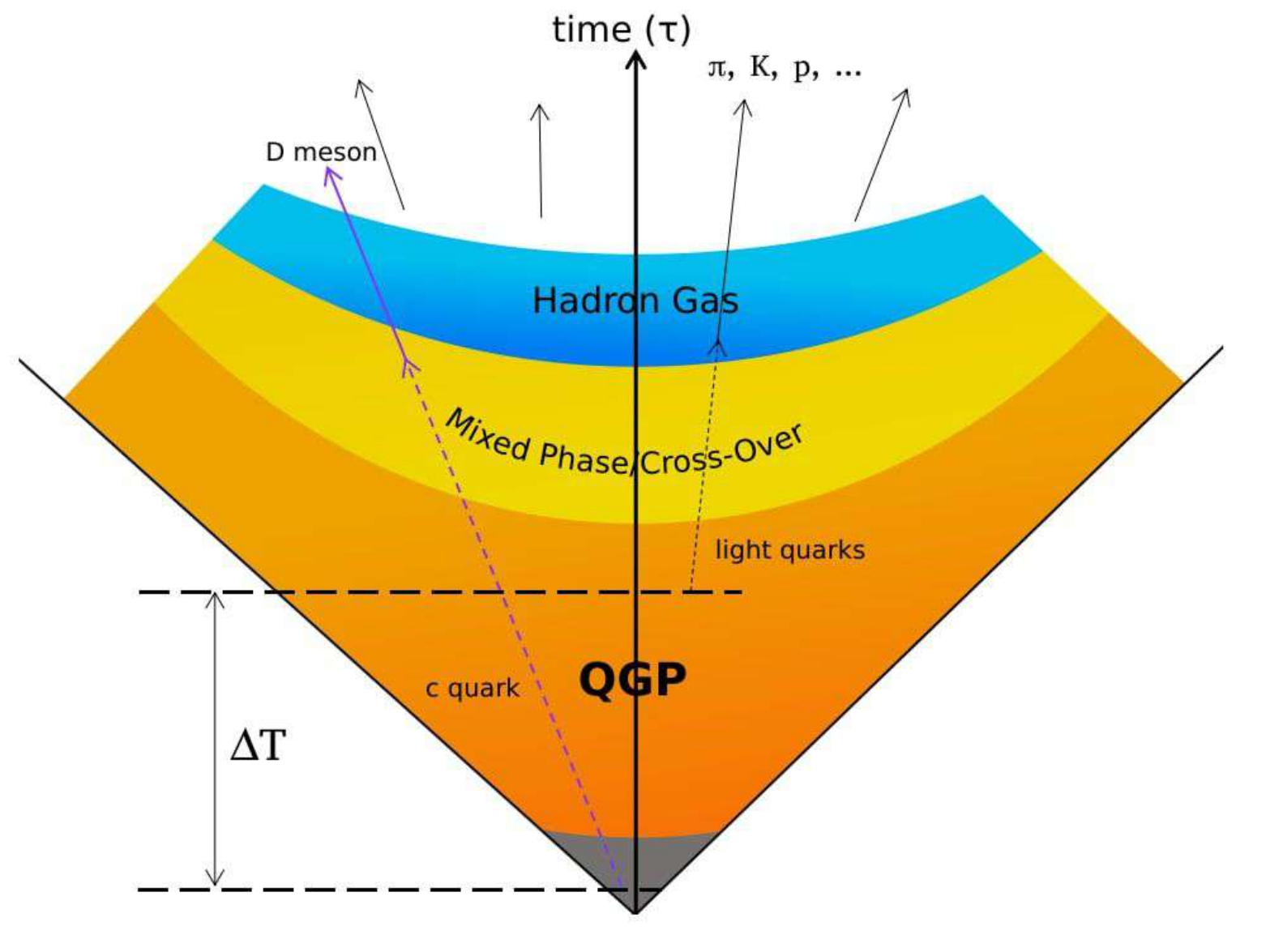}
  \caption{Schematic view of the evolution of matter in a high-energy nucleus-nucleus collision. Heavy $c$ quarks are produced in the initial stages, while light quarks are formed in the QGP phase. Later, during the phase transition, quarks combine to form hadrons.}
  \label{fig:cinGQP}
\end{figure}

In a high-energy collision, light-flavour and strange hadrons\footnote{From here on we refer to them simply as light-flavour hadrons.} mostly carry information about the final state~\cite{Andronic:2017pug}. By detecting heavy-flavour hadrons, however, one can learn about the earlier stages of the reaction as well. The heavy-flavour charm ($c$) and beauty ($b$) quarks are produced in the initial stages of the collisions. Since they have a very long lifetime and negligible annihilation cross section, they survive the collision and experience the whole evolution of the system~\cite{Andronic:2015wma, Sahoo:2022iul}. 
In pp collisions they serve as a testbench for QCD, since the processes can be perturbatively described down to low momenta. In heavy-ion collisions, however, they come into contact with the QGP through collisional and radiative interactions before forming hadrons. They are not expected to fully thermalize, and they bring information about the whole evolution of the QGP (Fig.~\ref{fig:cinGQP}). Recent development of detector capabilities allows for high-precision measurements of heavy-flavour hadrons over a broad range of transverse momentum (\pT) down to fractions of a GeV/$c$. The most common heavy-flavour hadrons, emerging from open charm processes, are the D$^0$, D$^+$ and D$^{*+}$ mesons consisting of a $c$ quark and a light antiquark, and their antiparticles (collectively referred to as D mesons). The identified spectra of D mesons were measured at RHIC and LHC energies in several collisional systems with high precision, which sets the stage for a thorough comparison of light and heavy-flavour hadron production at the next upgrade phase of the ALICE detector~\cite{ITS3} or the proposed ALICE~3~\cite{ALICE:2022wwr}. 

Spectra of identified particles produced in high-energy collisions comprise soft thermally produced particles, as well as particles from hard QCD processes. The non-extensive Tsallis\,--\,Pareto statistical framework describes these two components of the spectra in a unified way, independently of their formation process~\cite{Tsallis:1987eu, Tsallis:2009tri, Tsallis:2009zex}. The Tsallis-thermometer proved to be a sensitive tool to learn about the system size and/or fluctuations and correlations in the system. This was proposed in Ref.~\cite{Biro:2020kve}. However, the method had been previously applied only to light-flavour hadrons, therefore the results are mostly representative of the final state. In the current work, we evaluate the applicability of non-extensive thermodynamical principles on heavy-flavor production, and use D mesons to investigate the thermodynamical properties of earlier stages of the system.

\section{Method}

The transverse momentum distributions of charged and identified hadrons, resulting from energetic collisions of hadrons or heavy nuclei, offer valuable insights into the particle creation processes and the properties of QCD matter that may be produced in these reactions. The soft, low-\pT part of the spectrum is associated with the particles stemming from a thermal equilibrium. This part is well described by Boltzmann\,--\,Gibbs statistics, characterized by the kinetic freeze-out temperature parametrizing the exponential function~\cite{Cleymans:1992zc, Braun-Munzinger:2003pwq}. On the other hand, the jetty, high-\pT regime of the spectrum follows a power-law-like distribution, inherited from perturbative QCD hadron production~\cite{Wong:2015mba}.

The Tsallis\,--\,Pareto family of distributions, derived from the generalization of the traditional Boltzmann\,--\,Gibbs entropy, successfully unite the two regions of the spectrum~\cite{Tsallis:1987eu, Tsallis:2009tri, Tsallis:2009zex}. 
It has been shown in several recent studies that various parametrizations of these distributions fit well to the measured data from RHIC to LHC energies~\cite{Tsallis:1987eu, Cleymans:2012ya, Cleymans:2013rfq, Wong:2015mba, Shen:2019kil, Cleymans:2020ojr, Yang:2021bab, Biro:2020kve, Gu:2022xjn, Chen:2023zbf, Badshah:2023sfs}. 
While thermal equilibrium is often assumed implicitly in heavy-ion collisions, it is debated whether it can form~\cite{Badshah:2023sfs,Castorina:2019xkl}, and it is even less likely to be encountered in small collision systems. Fortunately, this assumption is not necessary for the understanding of the observed patterns of the final state~\cite{Biro:2003vz}. The non-extensive statistical framework can be applied to small as well as rapidly evolving systems, without requiring thermal equilibrium~\cite{Alqahtani:2022xvo}.
One of the motivations of our study is to extend the limits of applicability of thermodynamics in such systems, utilizing theoretical developments from the past decade.

In the present work, we consider a thermodynamically-motivated and consistent form of the distribution to fit the invariant yields of identified hadrons at mid-rapidity:
%
\begin{eqnarray}
  \left.\frac{\dd^2N}{2\pi \pT \dd \pT \dd y}\right|_{y\approx0} \equiv &  
    A m_T f^q =  & A \mT \left[1+\frac{q-1}{T}(\mT-\mu) \right]^{-\frac{q}{q-1}},
  \label{eq:TS}
\end{eqnarray}
%
where $A= gV/(2\pi)^3$ is the normalization factor containing the volume $V$ and degeneracy factor $g$; $q$ and $T$ are the non-extensivity parameter and the Tsallis temperature, respectively; $\mT=\sqrt{\pT^2+m^2}$ is the transverse mass, and the chemical potential $\mu$ is approximated as the rest mass of the given hadron, $\mu\approx m$. 
This is a theoretically motivated~\cite{Letessier:2002ony,Cleymans:2020ojr} pragmatic choice supported by data-driven studies~\cite{Shen:2019kil}.
Following the works of \cite{e16126497, Biro:2014yoa, Biro:2017arf, Biro:2020kve}, we can define the Tsallis parameters from the event-by-event fluctuations of the number of the produced particles, $n$:
\begin{eqnarray}
  T &= &\frac{E}{\left<n\right>},  \label{eq:fluctT}\\
  q &= &1-\frac{1}{\left<n\right>}+\frac{\Delta n^2}{\left<n\right>^2}. \label{eq:fluctq}
\end{eqnarray}
This leads to an energy-dependent linear correlation between the Tsallis parameters 
\begin{equation}
    T=E(\delta^2-(q-1)),
    \label{eq:Edelta2}
\end{equation}
where the relative size of multiplicity fluctuations $\delta^2 := \frac{\Delta n^2}{\left<n\right>^2}$ is assumed to be constant. Note that in case of flat enough rapidity distributions at mid-rapidity, the $y\approx 0$ limit can be used and $E$ reduces to $m_{\rm T}$.

In~\cite{Biro:2020kve} we have shown that the Tsallis parameters present a strong correlation with the average event multiplicity, spanning several collision systems. For identified light-flavour hadrons the Tsallis parameters also exhibit a scaling behaviour in charged-hadron multiplicity ($\Nch \sim n$), as well as in collision energy (\sqsn). The scaling results in a grouping of the Tsallis parameters $T$ and $q$ towards low event multiplicities, around the values $T_{\rm eq}=0.144\pm0.010$ GeV and $q_{\rm eq}=1.156\pm0.007$. 
However, as the production time-scale (and consequently, the carried information about interacting material) of heavy-flavour hadrons is different, it is an interesting opportunity to quantitatively examine these hadron species with respect to the Tsallis-thermometer.

\section{Results and Discussion}
\label{sec:expdata}

In the current study, we analyzed 11 datasets of pp, p--Pb, Au--Au and Pb--Pb collision systems at various center-of-mass energies from the publicly available measurements of ALICE and STAR experiments.
From the pp collisions at $\sqs = 5.02$~TeV and 7~TeV energies from the ALICE experiment we investigated the minimum-bias spectra of D$^0$, D$^+$, and D$^{*+}$ mesons~\cite{ALICE:2017olh,ALICE:2019nxm}. The spectra of the same three meson species were also taken from the p--Pb collisions at $\sqsn = 5.02$ TeV energy from  ALICE~\cite{ALICE:2019fhe}. We also studied two centrality-dependent (and consequently multiplicity-dependent) D$^0$-meson datasets of A--A systems. One of them was the Au--Au
system at $\sqsn = 200$ GeV energy from the STAR with a total of 5 centrality classes: 0--10\%, 10--20\%, 20--40\%, 40--60\%, 60--80\%~\cite{STAR:2018zdy}. The other one was Pb--Pb  at $\sqsn = 2.76$ TeV center-of-mass energy from the ALICE experiment with 0--20\% and 40--80\% centrality classes~\cite{ALICE:2012ab}. All the spectra cover a broad range of transverse momentum over a large number of data points, which allows for a precise application of the Tsallis\,--\,Pareto fit~\cite{Shen:2019kil}.

The investigated D-meson datasets were fitted with the Tsallis\,--\,Pareto distribution using the LMFIT: Non-Linear Least-Squares Minimization and Curve-Fitting for Python package~\cite{zenodo}. At first, the low- and high-$p_{\rm T}$ regions were fitted separately, after which the results of these fits were used as an input for the fit of the whole spectrum. This method proved to be stable and provided good $\chi^2/ndf$ values. The fit parameters are summarized in table~\ref{tab:fits}, while the fitted spectra are shown in~\ref{app:fits}.

The presence of radial flow may slightly alter the outcome of the fits. Studies that focused on the $\pT<3$ GeV/$c$ part of the spectrum that incorporated a blast wave model found non-zero radial flow in RHIC heavy-ion collisions, while for proton-proton collisions radial flow was found to be zero~\cite{Tang:2008ud}. In case of LHC energies, a substantial radial flow was found even in proton-proton collisions~\cite{Khuntia:2018znt,Jiang:2013gxa}. Note that the current description, which utilizes the non-extensive framework within a co-moving reference frame \cite{Urmossy:2009jf} and uses inputs from broader \pT ranges (from 0 up to 8--30 GeV/$c$ depending on the particle species), does not require a radial flow parameter to account for the final-state hadrons produced in heavy-ion collisions. However, radial flow is a complex phenomenon that can have experimental consequences not covered by a single blast-wave fit parameter. The precise extraction of the characteristic radial flow from the hadron spectra would require a more sophisticated calculation, which is beyond the scope of the current study.

\begin{table}[h]
    \caption{Parameters of Tsallis fits on D-meson spectra.}
    \label{tab:fits}
    \begin{indented}
    \lineup
    \item[]\begin{tabular}{@{}llllll}
    \br
      \sqsn (GeV) & Hadron & $\left<\frac{\rm d N_{ch}}{\rm d \eta}\right>$ & $T$ (GeV) & $\0\0\0\0q$ & $\chi^2/ndf$\\
    \mr
    \multirow{5}{*}{AuAu, 200} & \multirow{5}{*}{D$^0$} & 680.0 & 0.32$\pm$0.01& 1.06$\pm$0.01& \0\04.1/8\\
    & & 424.5 &0.36$\pm$0.03& 1.05$\pm$0.02& \015.6/8 \\
    & & 235.7 & 0.31$\pm$0.01& 1.07$\pm$0.01& \012.7/8\\
    & & \090.0 & 0.32$\pm$0.02& 1.09$\pm$0.01& \013.8/8\\
    & & \027.0 & 0.29$\pm$0.04& 1.12$\pm$0.03& \029.2/8\\
    \multirow{2}{*}{PbPb, 2760} & \multirow{2}{*}{D$^0$} & 600.0&0.32$\pm$0.06 & 1.16$\pm$0.02& \0\00.9/4\\
    & &\045.0 & 0.23$\pm$0.05& 1.21$\pm$0.02& \0\00.9/4\\
    \multirow{3}{*}{pp, 5020} & D$^0$ & & 0.45$\pm$0.01&1.16$\pm$0.01 & \0\07.8/18\\
    & D$^+$ &  & 0.43$\pm$0.02&1.16$\pm$0.01 & \015.9/17\\
    & D$^{*+}$ &  & 0.44$\pm$0.02& 1.17$\pm$0.01& \011.6/16\\
    \multirow{3}{*}{pPb, 5020} & D$^0$ & & 0.43$\pm$0.02& 1.17$\pm$0.01& 199.7/18\\
    & D$^+$ &  & 0.38$\pm$0.03& 1.17$\pm$0.01& 792.3/17\\
    & D$^{*+}$ &  & 0.42$\pm$0.02& 1.17$\pm$0.01& \093.2/16\\
    \multirow{3}{*}{pp, 7000} & D$^0$ & & 0.49$\pm$0.02& 1.15$\pm$0.01&\0\07.5/10 \\
    & D$^+$ &  & 0.47$\pm$0.04& 1.16$\pm$0.01& \0\07.8/8\\
    & D$^{*+}$ &  & 0.44$\pm$0.04& 1.17$\pm$0.01& \0\07.5/8\\
    \br
    \end{tabular}
    \end{indented}
\end{table}

Following the method developed in~\cite{Biro:2020kve}, we evaluated the consistency of the thermodynamical variables~\cite{Cleymans:2012ya,Cleymans:1992zc} by verifying that the first law of thermodynamics with a first order Euler equation
\begin{equation}
    \epsilon + P - Ts - \mu n = 0
    \label{eq:consistency}
\end{equation}
is fulfilled.
We found that the consistency is fulfilled for light-flavour mesons within 1\% precision. For heavier hadrons this can deviate slightly more, however, for D mesons it always stays below 8\%. The details of this procedure are presented in~\ref{app:thermo}.

\begin{figure}[h]
  \centering
  \includegraphics[width=\linewidth]{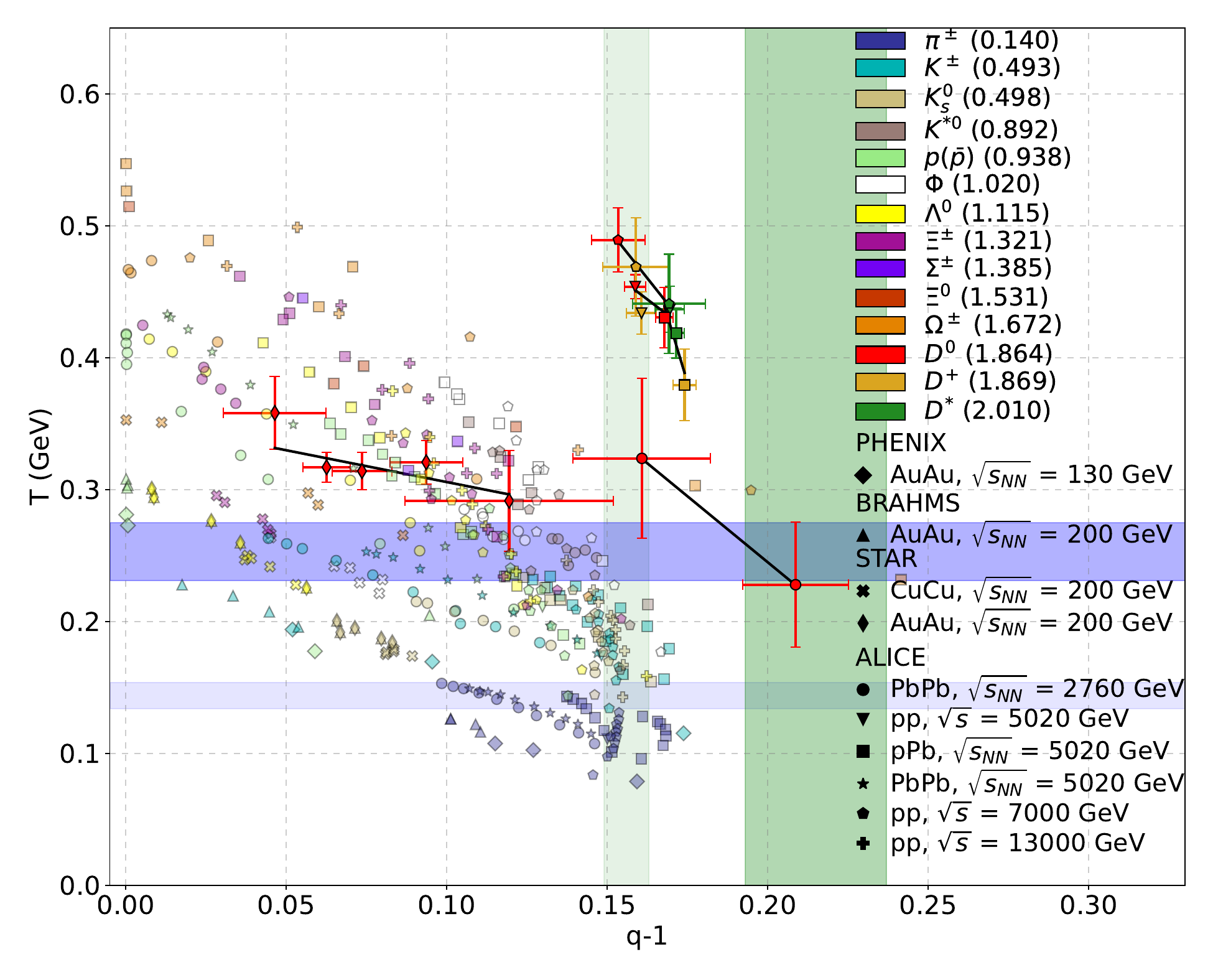}
  \caption{Tsallis-thermometer: the fitted $T$ and $q$ parameters of identified D$^0$, D$^+$ and D$^{*+}$ mesons stemming from pp, p--Pb, Pb--Pb, and Au--Au collisions at various energies and various multiplicity classes. Results for light-flavour hadrons from \cite{Biro:2020kve} are shown in semi-transparent colours for comparison.}
  \label{fig:Tvsq}
\end{figure}

The $T$ and $q$ parameters, extracted from the D-meson fits, are presented in the Tsallis-thermometer, $T$\,---\,$(q-1)$ diagram shown in figure~\ref{fig:Tvsq} alongside the parameters from the light-flavour study~\cite{Biro:2020kve}, which are drawn without error bars and with semi-transparent colours for better visibility.
The effect of radial flow on the Tsallis parameter $T$ has been estimated to be minute~\cite{Biro:2020kve}, therefore we do not apply a correction for it.

For the case of light-flavour hadrons, a mass hierarchy had been observed: the parameter $T$ shows an increasing trend both with particle mass and with event multiplicity within the same particle species, while the parameter $q$ is decreasing in the same manner. This behaviour becomes more distinct for heavy hadrons, and is similarly observed for the D mesons. Strong dependence on multiplicity for heavy-flavour hadrons is also present.

Nevertheless, there are differences between the values of the Tsallis parameters corresponding to heavy- and light-flavour hadrons. This can give us an insight into the different production mechanisms and timescales. D mesons show a grouping based on the center-of-mass energy and collision system that is stronger than seen in light flavour hadrons with similar masses. D mesons from $\sqsn=200$ GeV Au--Au collisions are located towards the bottom left of the $T$\,---\,$(q-1)$ diagram, while the points of $\sqs=5.02$~TeV and 7~TeV collisions can be found towards the top right part of the diagram. This may be explained by the fact that in small systems, $c$ quarks come directly from the early stages of the collisions, corresponding to high $T$ values as well as larger non-extensivity parameter $q$. Furthermore, in A--A collisions $c$ quarks may experience coalescence within the cooling and expanding medium~\cite{Plumari:2017ntm}, therefore showing smaller Tsallis parameter values.

Another feature of the Tsallis-thermometer of light-flavour hadrons was the grouping of all the hadrons at small multiplicities around specific common $T_{\rm eq}$ and $q_{\rm eq}$ values. For the D mesons, the behaviour is similar, however, both the common Tsallis parameter values are higher. To determine these, we used the correlation~(\ref{eq:Edelta2}) to fit different sets of D meson data. Since D meson species have similar masses (1.864 GeV for D$^0$, 1.869 GeV for D$^+$ and 2.010 GeV for D$^{*+}$), and data are limited, we considered them as one 'general type' D meson here. Each of the collision systems and energies (pp 5.02 TeV, pp 7 TeV and p--Pb 5.02 TeV) were used in this sense, by providing enough high accuracy for our conclusions.

\begin{figure}[t]
  \centering
  \includegraphics[width=\linewidth]{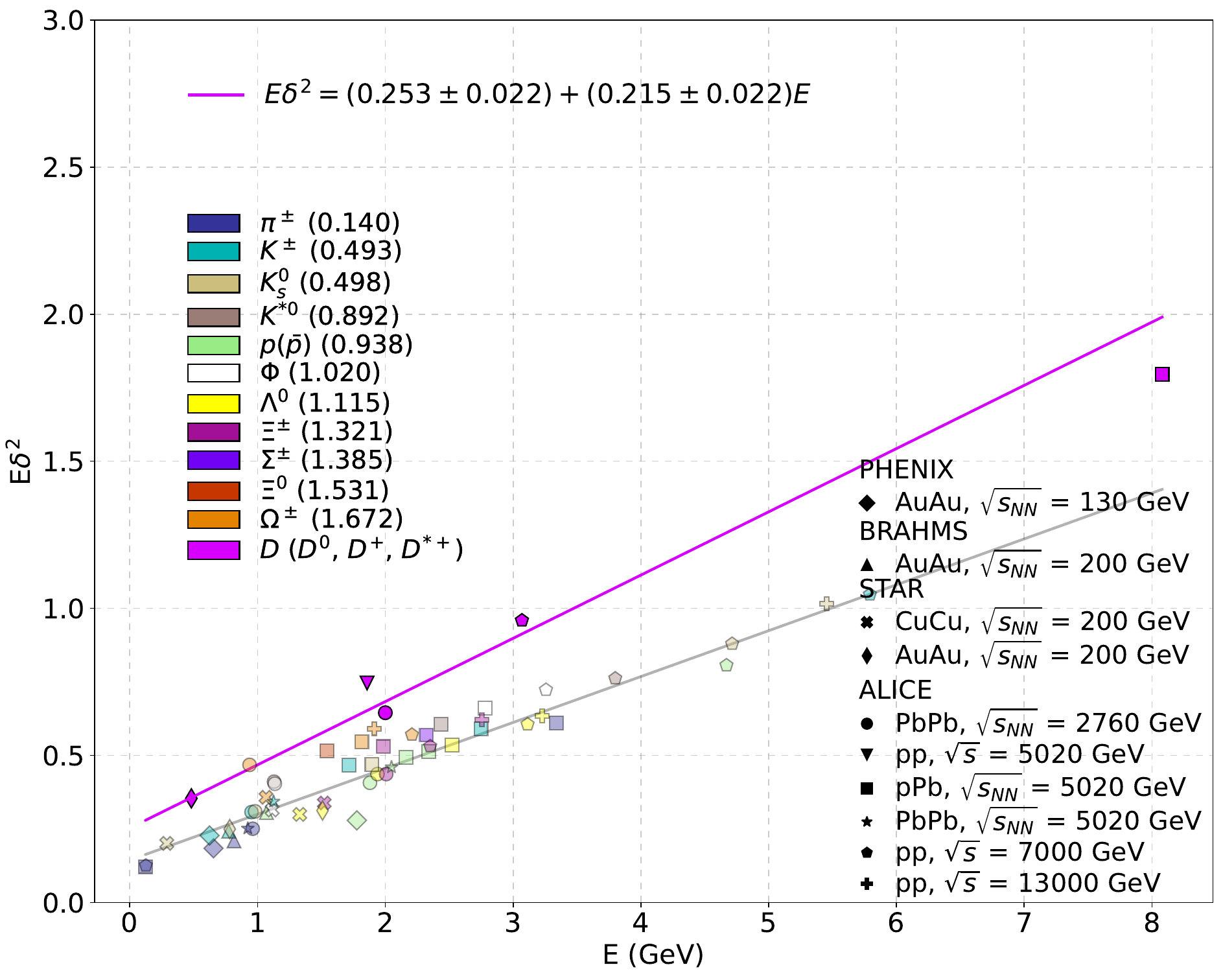}
  \caption{\label{subfig:Evsd2}Correlation of the fit parameters $E\delta^2$ as a function of $E$. D-meson results (magenta) are compared to light-flavour hadron results (semi-transparent) from~\cite{Biro:2020kve}.}
\end{figure}

The fitted correlation between the parameters $E$ and $\delta^2$ are presented in the $E$---$E\delta^2$ diagram in Fig.~\ref{subfig:Evsd2}, where energy is associated with the transverse mass $m_{\rm T}$ at mid-rapidity. The points from fits to the D mesons are compared to those from fits to the light-flavour hadrons, with the uncertainties omitted from the figure for better visibility. One can observe that the relative size of the multiplicity fluctuations $\delta^2$ is larger for all D meson points compared to the light-flavour hadrons. Furthermore, the points from all fits to small systems correspond to larger $E$ values than those from heavy-ion collisions, as the fit determined by these points in the $T$---$(q-1)$ diagram is steeper.

To find the values of the common Tsallis parameters for D mesons, we inverted~(\ref{eq:Edelta2}) to the form of
\begin{equation}
    E\delta^2=T-(q-1)E \ . 
\end{equation}
After fitting D-meson points with this equation we got the values $T_{\rm eq}=0.253 \pm 0.022$~GeV and $q_{\rm eq}=1.215\pm0.022$. These values together with their uncertainties are shown in Fig.~\ref{fig:Tvsq} as the blue and the green band, respectively. Therefore, values of the common Tsallis parameters for D mesons are offset by $\Delta T_{\rm eq}=0.109\pm0.024$~GeV and $\Delta q_{\rm eq}=0.059\pm 0.023$ compared to the light hadrons. This difference, illustrated in Fig.~\ref{fig:cinGQP} as $\Delta T$, can be explained by the different information which is carried by light and heavy-flavour hadrons. Light flavour is predominantly formed during the kinetic freeze-out phase of a collision, while D mesons are formed by the c quarks, which originate in the early stages of the collisions. Therefore D mesons are probes of a much hotter medium, such as the QGP, which results in an increased Tsallis-temperature value. 

\begin{figure}[h]
  \centering
  \includegraphics[width=0.8\linewidth]{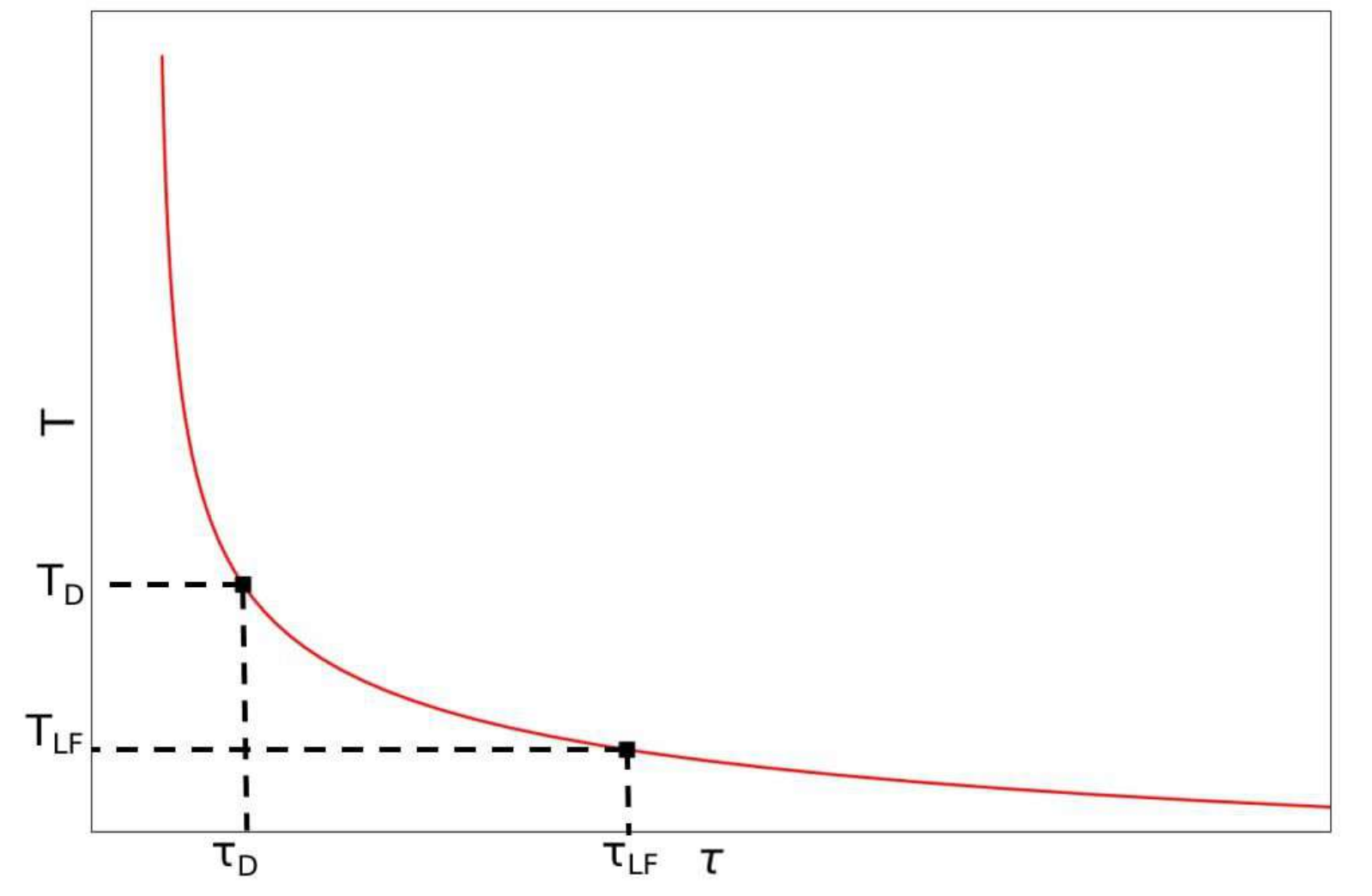}
  \caption{Schematic cooling curve of an ultrarelativistic collision of hadrons. Temperature and time scales are shown in arbitrary units.}
  \label{fig:TvsTau}
\end{figure}

The Tsallis temperature $T_{\rm eq}$ obtained separately for light-flavour hadrons and D mesons from the non-extensive thermodynamical model provide an opportunity to connect the initial proper time corresponding to charm creation, $\tau_{\rm D}$ to that of light flavour, $\tau_{\rm LF}$. In order to estimate the difference in the production time, a simple Bjorken model can be utilized as the expansion mechanism of the ideal, ultra-relativistic matter. Note that the Bjorken picture does not require any assumption on the thermodynamics, therefore the non-extensive framework can be applied to define a temperature-like measure. The Bjorken model gives the proper-time evolution of the energy density as $\varepsilon=3P=\sigma T^4$~\cite{Bjorken:1982qr, VOGT2007221}, and with the 
$\varepsilon(\tau_0)=\varepsilon_0\rightarrow T(\tau_0)= T_0$ initial condition, this has an analytic solution~\cite{Magas:2017nuw},
\begin{equation}
    \tau = \tau_0 \left( \frac{T_0}{T} \right)^3 \ .
\end{equation}
The applicability of the ultra-relativistic limit for the equation of state has been validated in Ref.~\cite{Biro:2020kve}. 
Assuming the same initial conditions for the light- and heavy-flavour hadrons, $T_{0,LF}(\tau_0)=T_{0,D}(\tau_0)$, the following ratio can be obtained as
\begin{equation}
    \tau_{\rm D} = \tau_{\rm LF} \left( \frac{T_{\rm LF}}{T_{\rm D}} \right)^3 \ .
\end{equation}
Substituting the appropriate $T_{\rm eq}$ values we get $\tau_{\rm D} = (0.18 \pm 0.06 ) \tau_{\rm LF}$, which shows that heavy-flavour hadrons bring thermodynamical information from significantly earlier times of the reaction than light-flavour hadrons as demonstrated in Fig.~\ref{fig:TvsTau}.

The Bjorken expansion can be calculated using a Tsallis-distribution utilizing a time-dependent $q$ parameter~\cite{Alqahtani:2022xvo}. Our result of $q_{\rm eq,D}>q_{\rm eq,LF}$ supports their conclusions on the decreasing value of the non-extensivity parameter with increasing $\tau$. In the current model, the change of the value, $\Delta q = 0.059 \pm 0.023$, represents a softening of the transverse momentum distribution between charm and light flavour, therefore it is a sensitive parameter of the time evolution as well. 
Using the current model we provided a numerical estimate for the ratio of light and heavy-flavor spectrum formation time scales. The relatively minor discrepancy in the trace anomaly might stem from the lower precision of the available heavy-flavor data or indicate the limits of the current framework. Further development of these models could be possible with more detailed and precise measurements extending to low \pT. Future high-precision detector systems with excellent secondary-vertex resolution capabilities, such as ALICE ITS3~\cite{ITS3} or the proposed ALICE3~\cite{ALICE:2022wwr} will offer excellent opportunities for such experiments.

\section{Summary}
\label{sec:summary}

In this paper we applied the Tsallis\,--\,Pareto non-extensive statistical framework to study the properties of D mesons. We showed that the transverse momentum distributions of heavy-flavoured D mesons are well described by the Tsallis\,--\,Pareto distribution, thus showing that it can be applied beyond the light flavours. The parameters from the fits fulfil thermodynamical consistency, therefore the statistical framework is applicable to D mesons as well. Similarly to the light-flavour case, the Tsallis parameters of the fits to D-meson data exhibit a scaling behaviour with charged particle multiplicity and with the collision energy. However, we showed that the scaling of D mesons is quantitatively different from that of light-flavour and strange hadrons. 

According to our hypothesis, $T_{\rm eq}$ and $q_{\rm eq}$ parameters were found to be higher for heavy-flavour than light-flavour hadrons. The higher $q_{\rm eq}$ means that the correlation within the heavy-quark system is slightly larger than that of light and strange quarks. This is due to the heavy quarks being produced in the early stages of a collision, where the volume of the system is smaller, while the energy density is higher. The $T_{\rm eq}$ parameter for D mesons is also higher. This can be explained with the similar reasoning, as coming from a much hotter state of the system, D mesons preserve this information, unlike the light-flavour hadrons. 

Considering an expanding system and the results from a non-extensive statistical approach, $\Delta T$ and $\Delta q$ on the Tsallis thermometer can be understood as time-frame projections of different stages of the time evolution. Comparing charm production to that of light flavour, we find that the production of D mesons corresponds to a significantly earlier proper time than light-flavour hadrons. Based on the Bjorken expansion, we estimated it as $\tau_{\rm D} = (0.18 \pm 0.06) \tau_{\rm LF}$.

\ack
This work has been supported by the NKFIH grants OTKA FK131979 and K135515, as well as by the 2021-4.1.2-NEMZ\_KI-2024-00031, 2021-4.1.2-NEMZ\_KI-2024-00033 and 2021-4.1.2-NEMZ\_KI-2024-00034 projects. The authors acknowledge the research infrastructure provided by the Hungarian Research Network (HUN-REN) and the Wigner Scientific Computing Laboratory.

\appendix
\section{\label{app:fits}Fits on D spectra}

\begin{figure}[ht]
  \centering   
\includegraphics[width=0.32\linewidth]{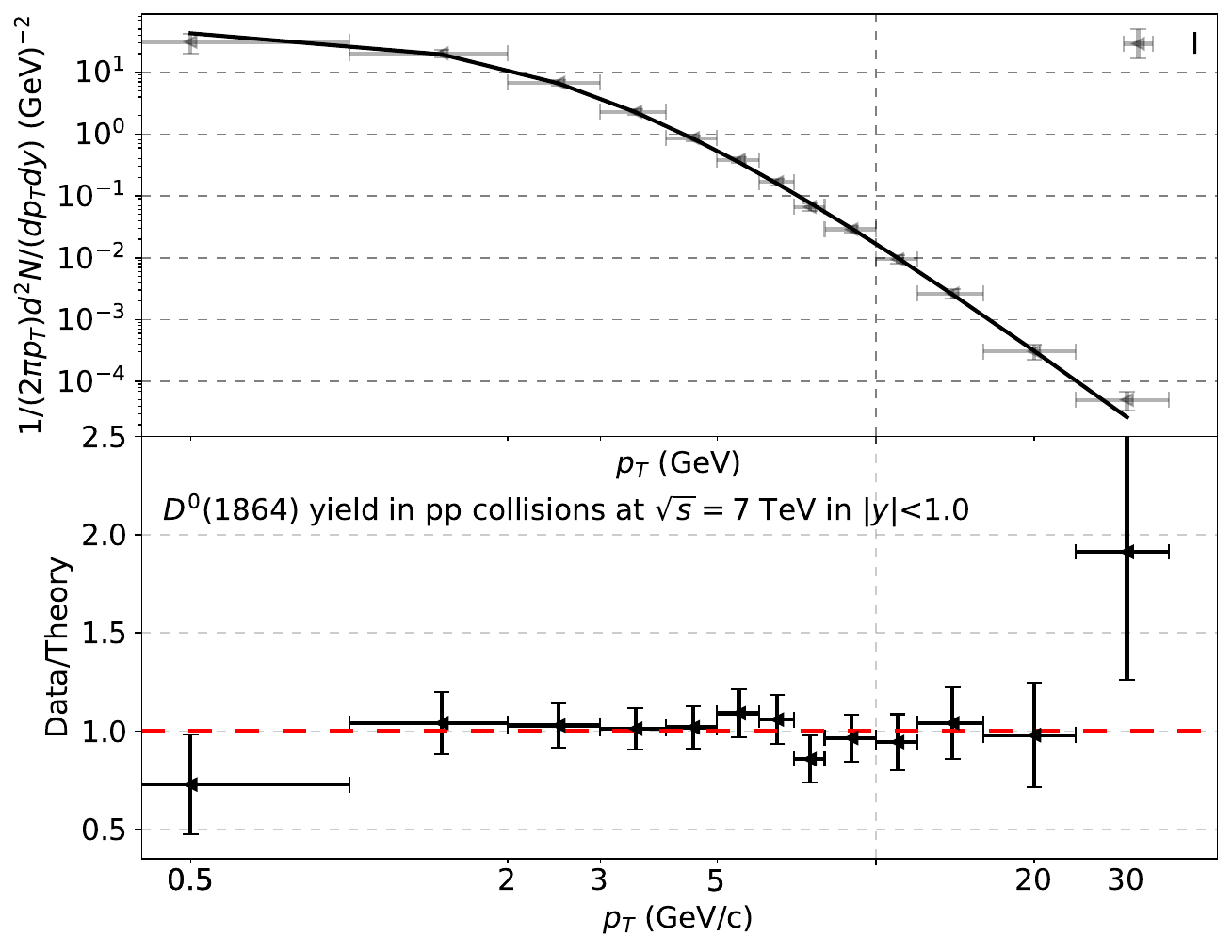}
\includegraphics[width=0.32\linewidth]{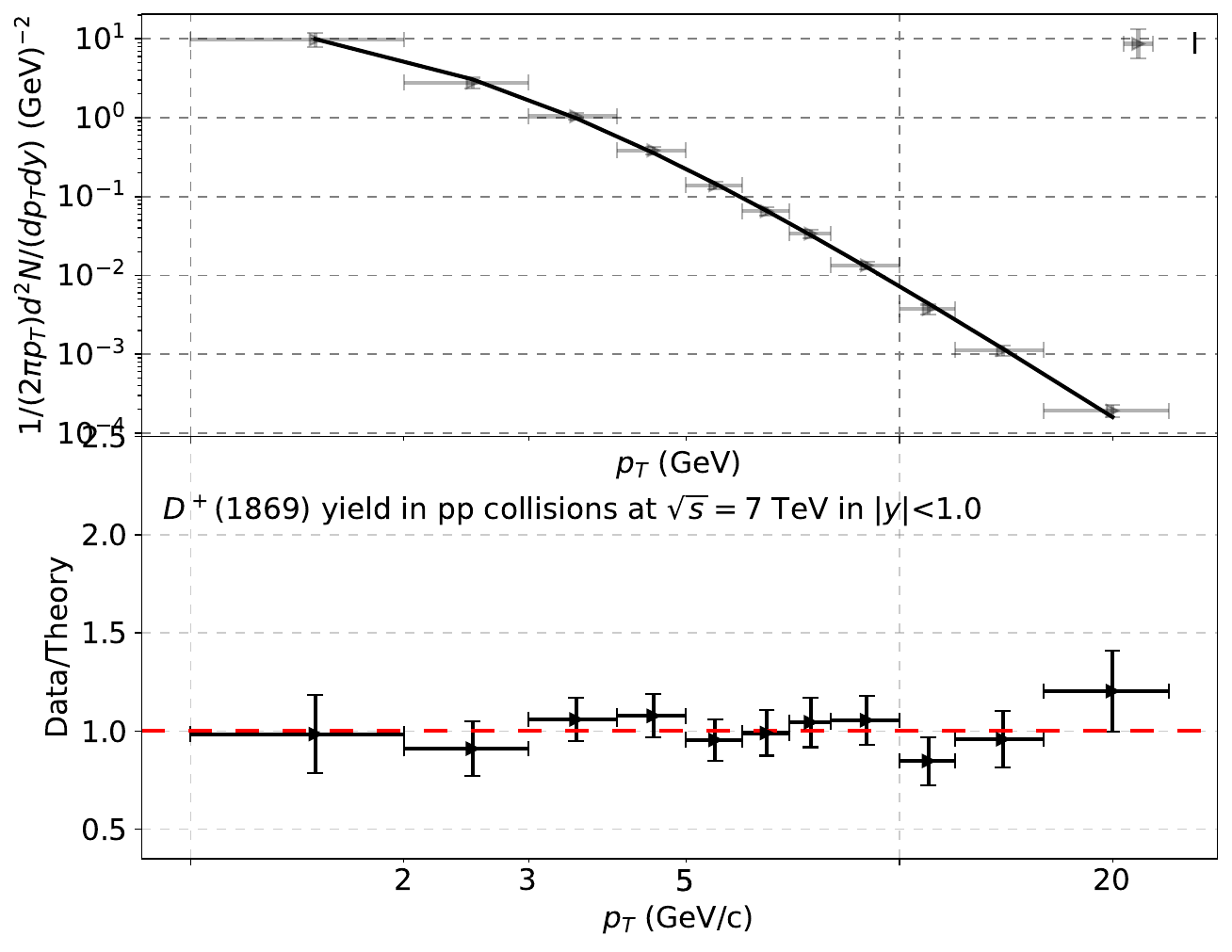}
\includegraphics[width=0.32\linewidth]{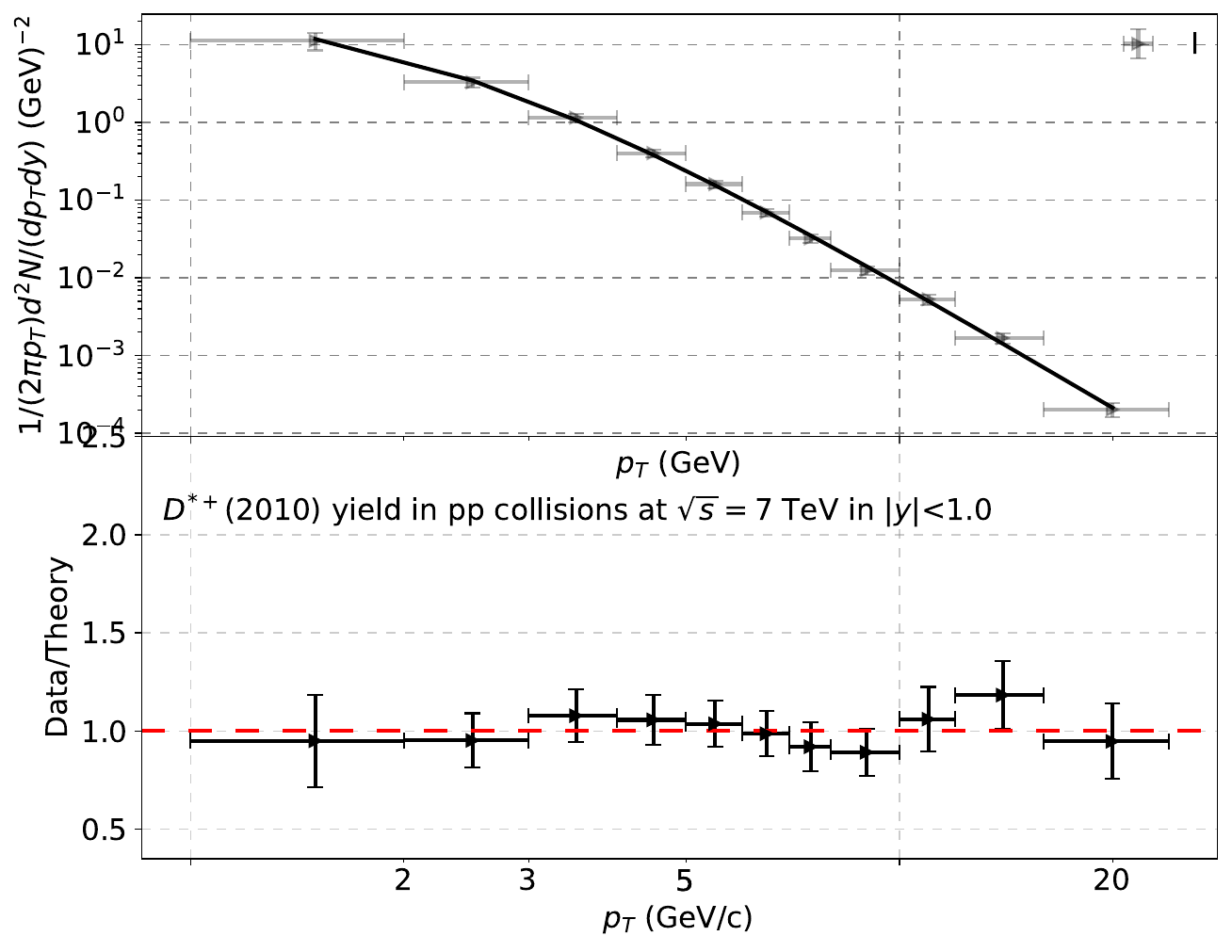}
\includegraphics[width=0.32\linewidth]{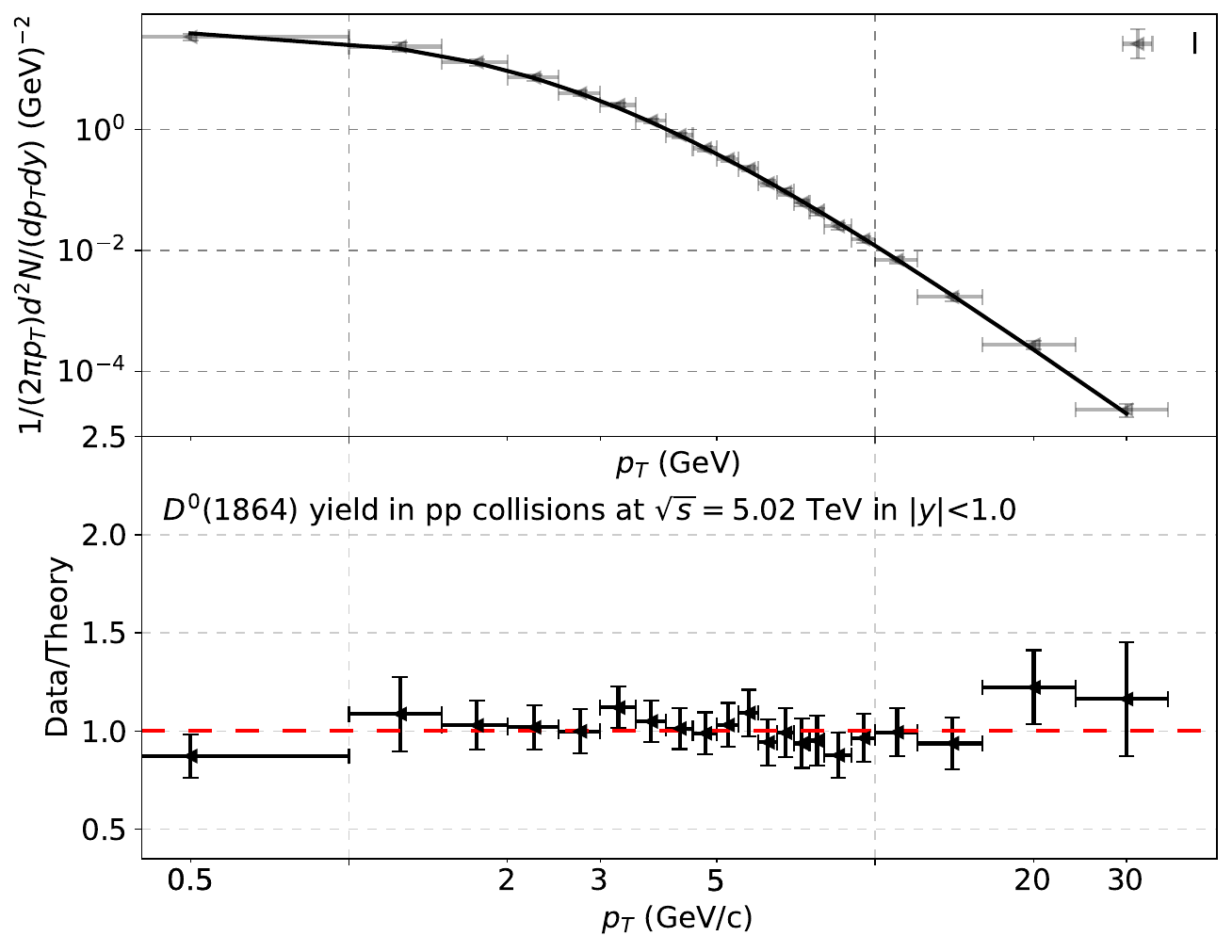}
\includegraphics[width=0.32\linewidth]{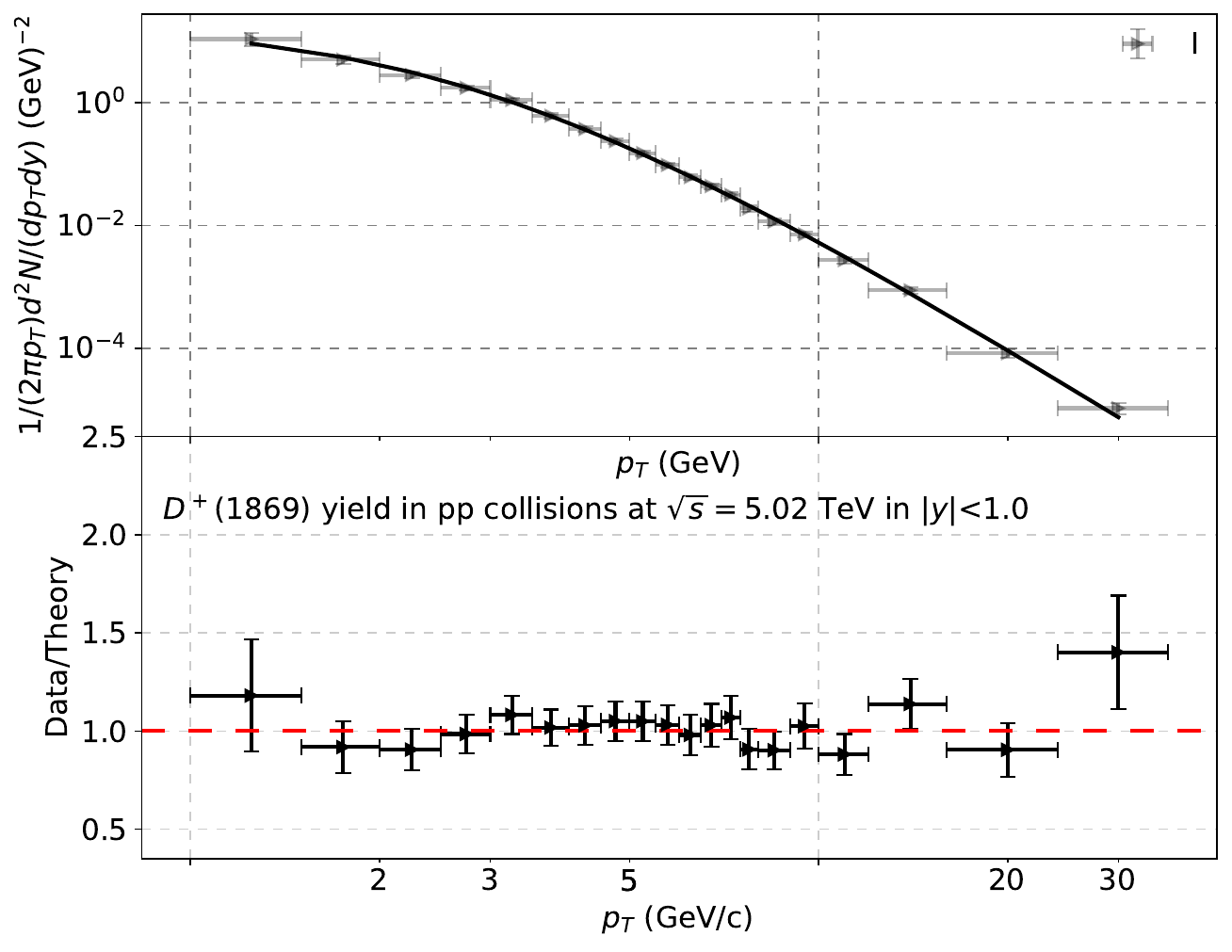}
\includegraphics[width=0.32\linewidth]{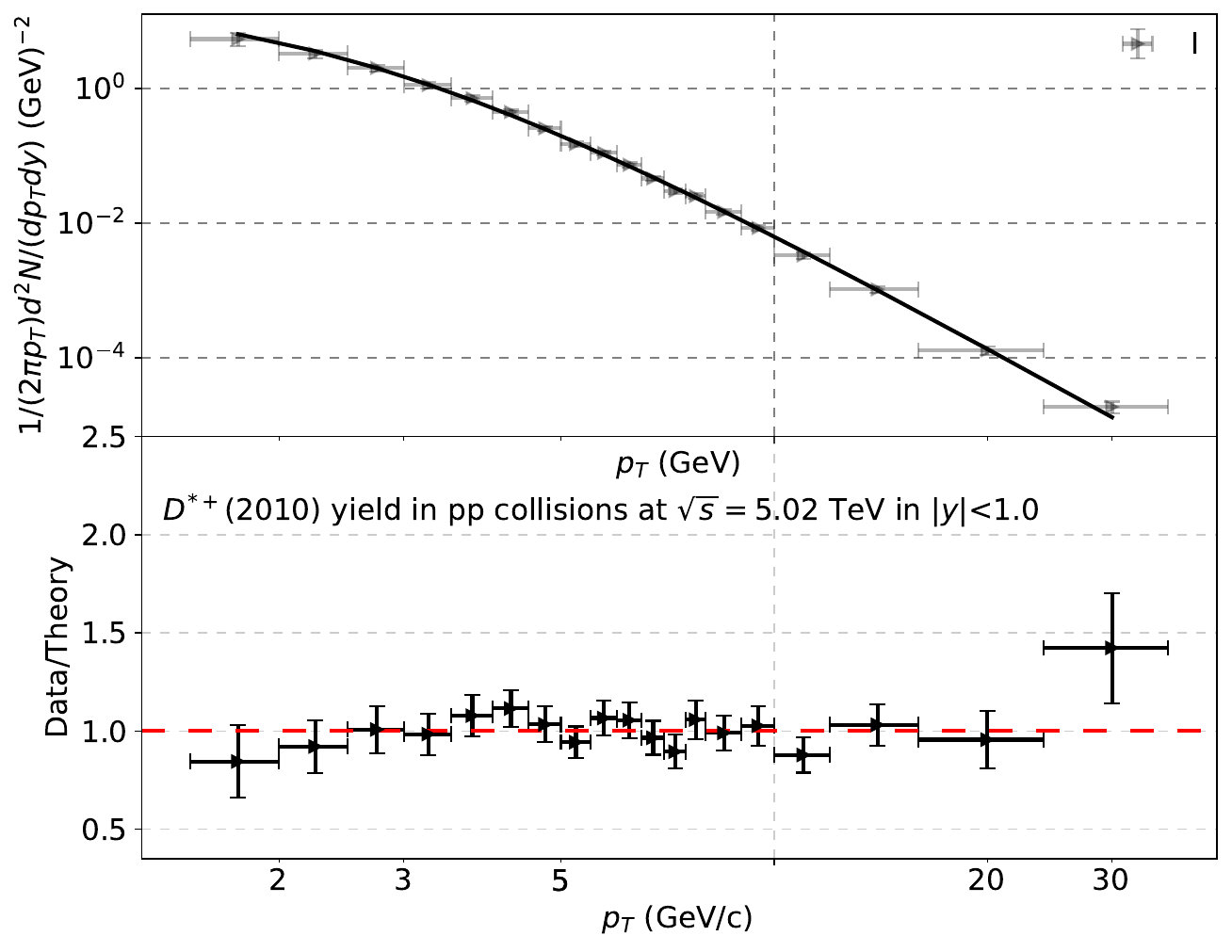}
\includegraphics[width=0.32\linewidth]{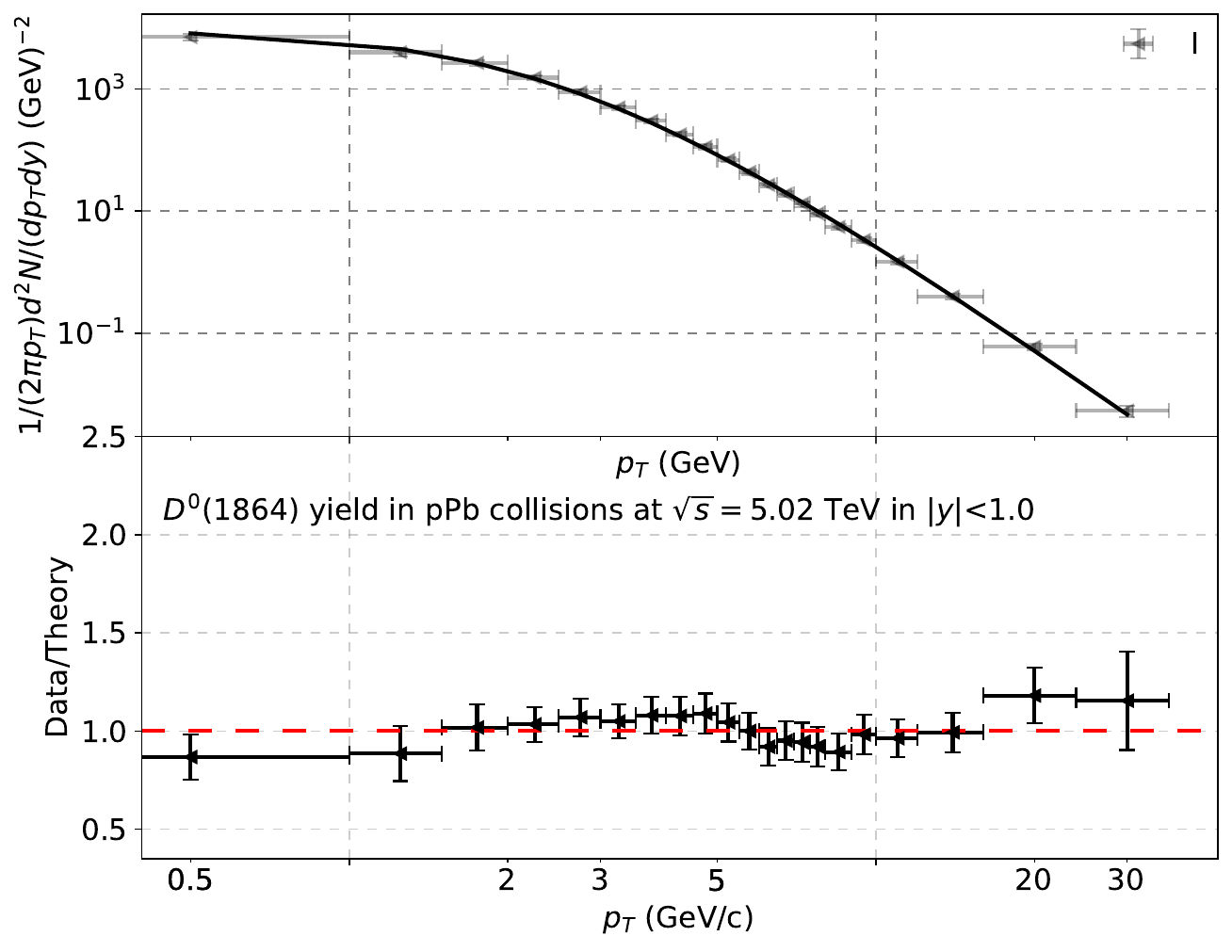}
\includegraphics[width=0.32\linewidth]{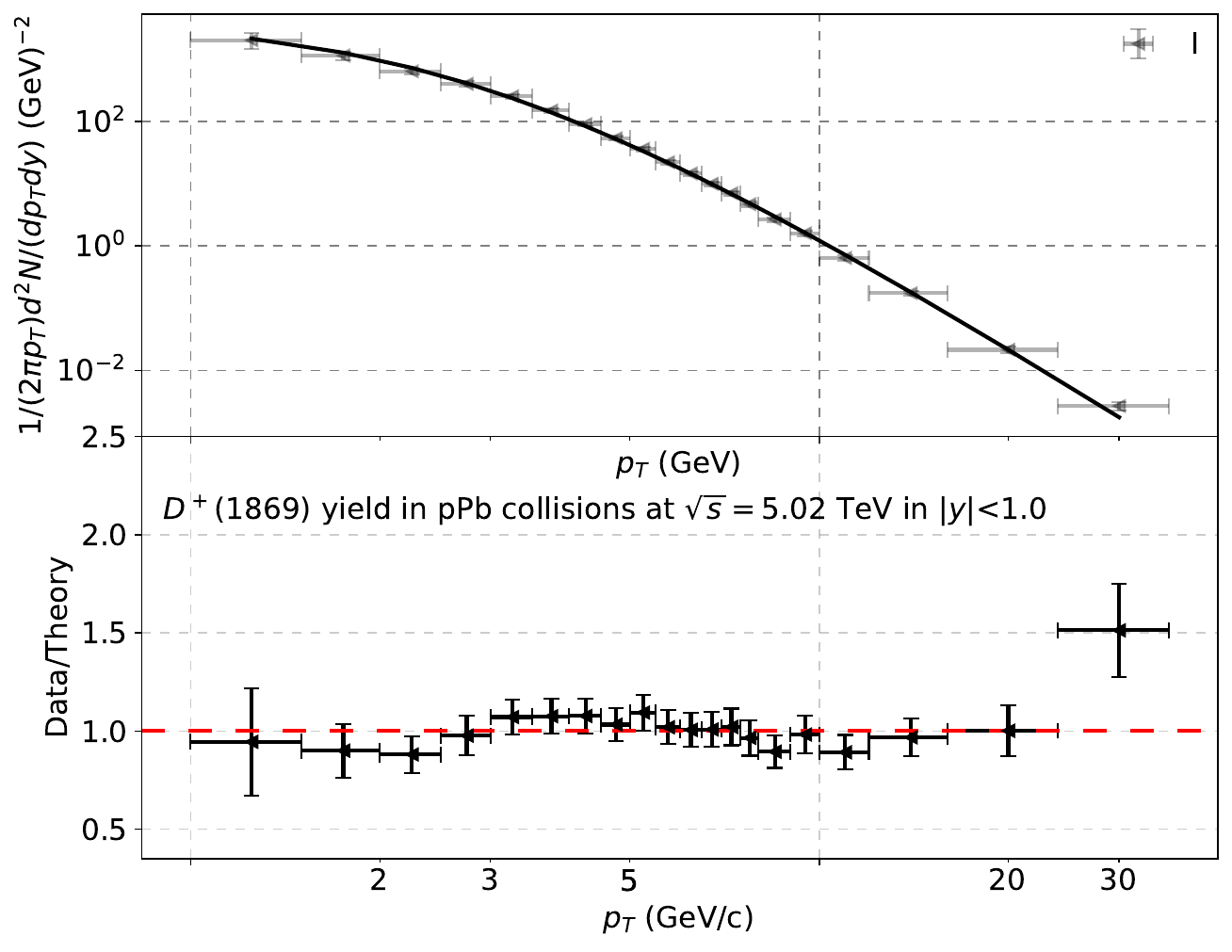}
\includegraphics[width=0.32\linewidth]{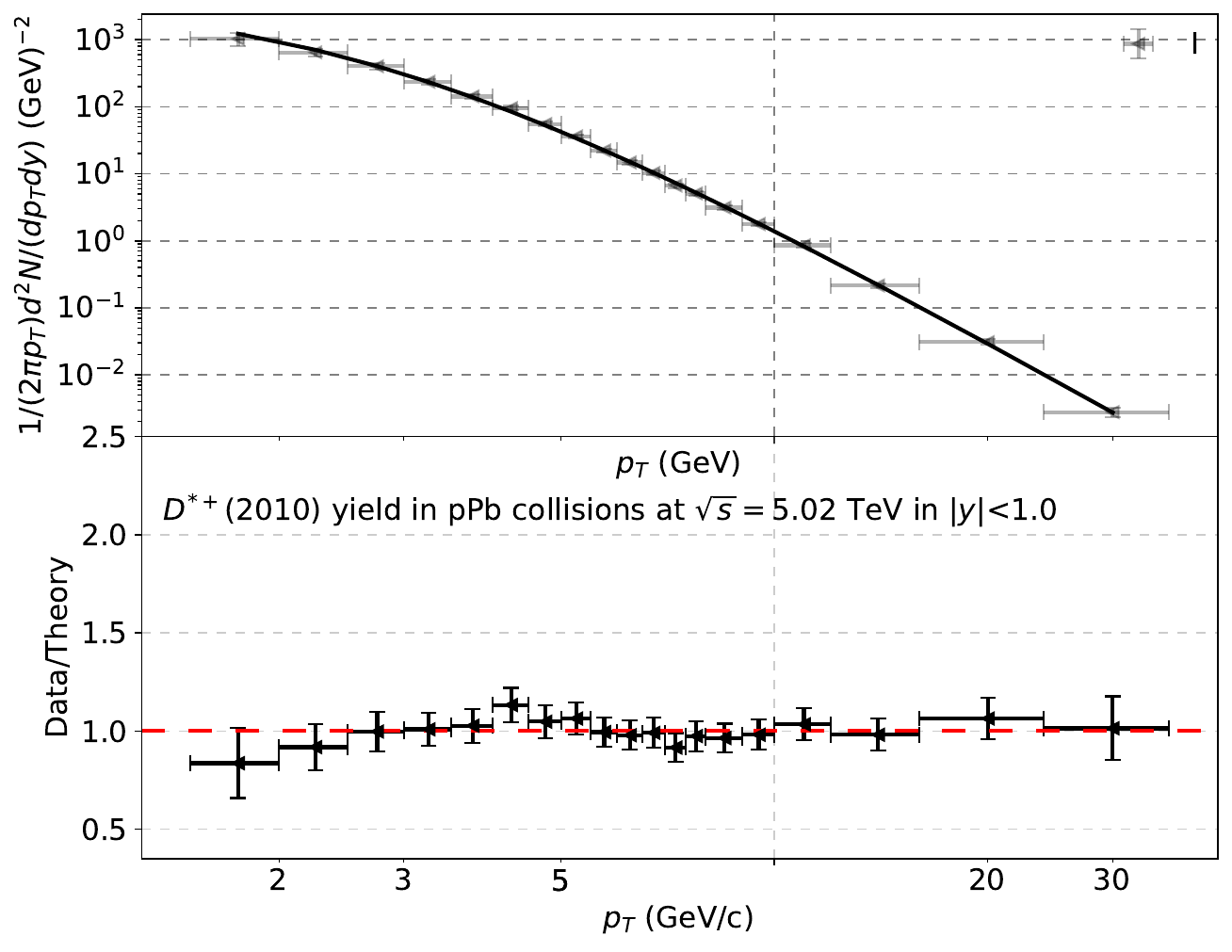}
\includegraphics[width=0.32\linewidth]{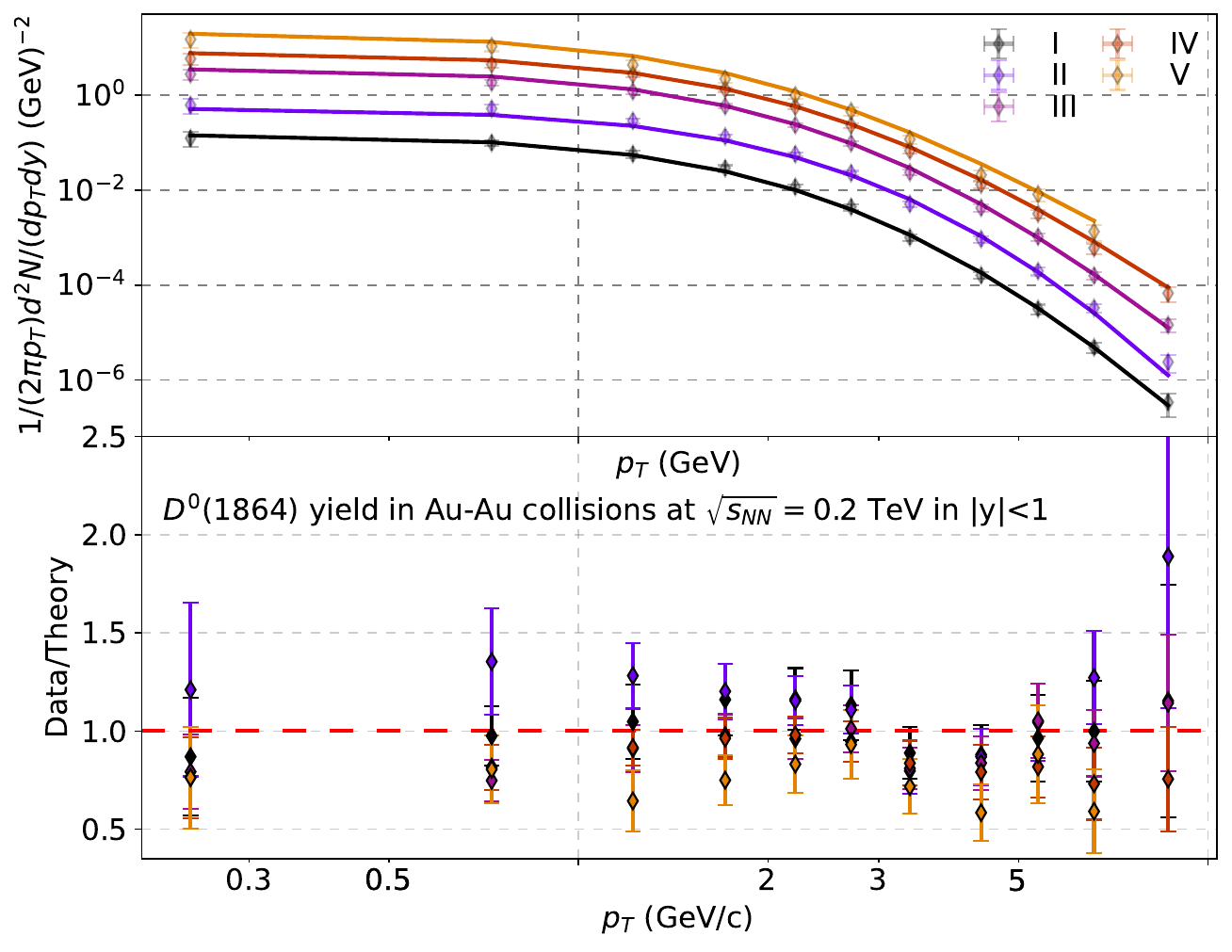}
\includegraphics[width=0.32\linewidth]{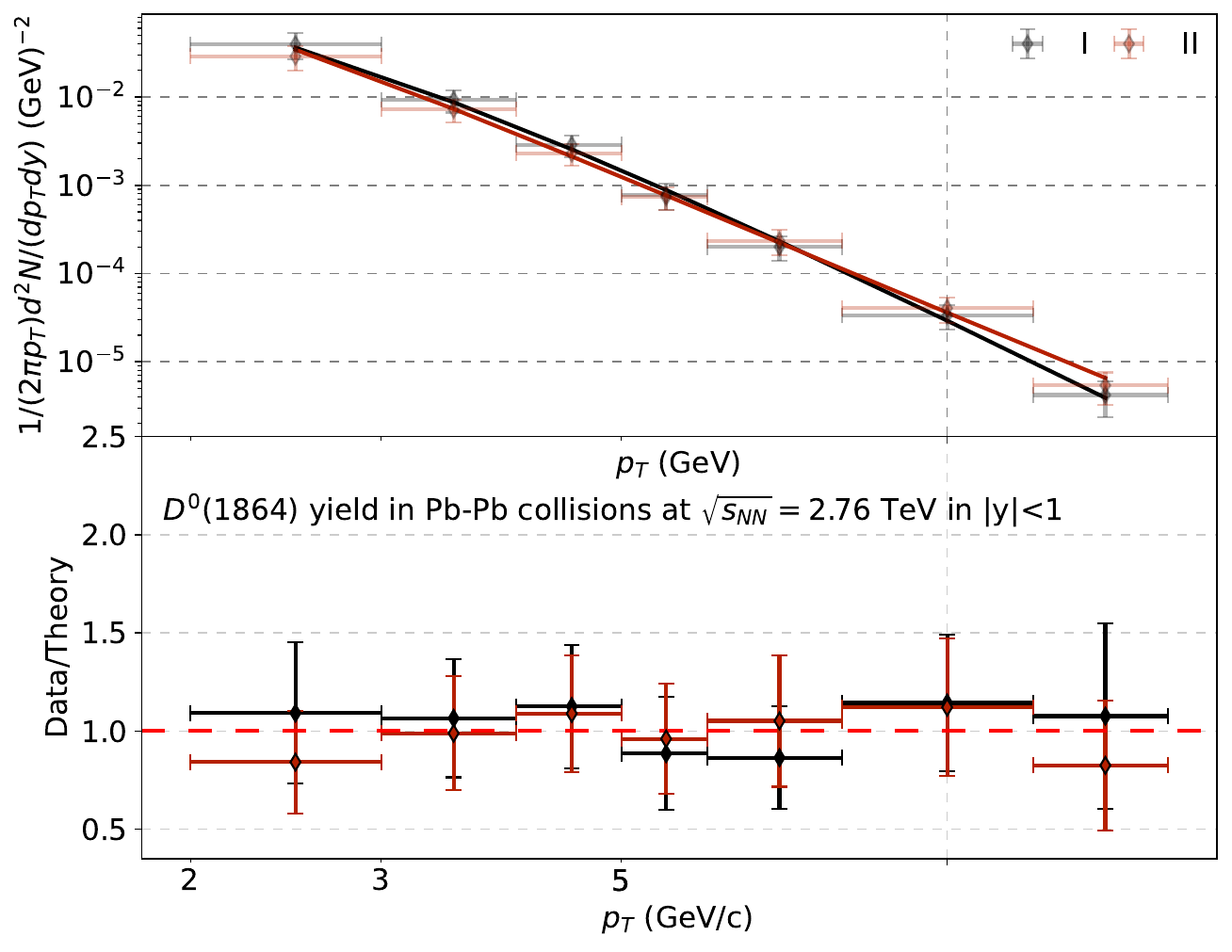}
\caption{Fits on the D meson spectra. The upper three rows from top to bottom correspond to ALICE pp data at $\sqs = 7$ TeV~\cite{ALICE:2017olh}, pp collisions at $\sqs = 5.02$ TeV~\cite{ALICE:2019nxm}, p--Pb data at $\sqsn = 5.02$ TeV~\cite{ALICE:2019fhe}. The lowest row shows STAR Au--Au data at $\sqsn = 200$ GeV (left)~\cite{STAR:2018zdy} and ALICE Pb--Pb data at $\sqsn = 2.76$ TeV (right)~\cite{ALICE:2012ab}.}
\label{fig:fits}
\end{figure}

Figure~\ref{fig:fits} shows the Tsallis fits for D-meson spectra from~\cite{ALICE:2012ab, ALICE:2019nxm, ALICE:2019fhe, ALICE:2017olh, STAR:2018zdy}.

\section{Thermodynamical consistency\label{app:thermo}}

The approach we use has been constructed according to first principles, therefore the model itself, including the derived Tsallis-Pareto distributions, satisfies eq.~\eref{eq:consistency}~\cite{Cleymans:2013rfq}. However, since the formula is applied on measured and derived observables, the consistency on measured data is granted only if the model can be applied and the quality of the data and the fits are good. The quantities in eq.~\eref{eq:consistency} are calculated from eq.~\eref{eq:TS} as follows~\cite{Cleymans:2013rfq,Cleymans:2012ya}, using the fitted Tsallis parameters, for a one component system:
\begin{eqnarray}
  P &= & g \int \frac{\dd^3p}{(2\pi)^3}  T f, \label{eq:P} \\
  N &= &nV = gV \int \frac{\dd^3p}{(2\pi)^3} f^q, \label{eq:N} \\
  s &= & g \int \frac{\dd^3p}{(2\pi)^3} \left[\frac{m_{\rm T}-m}{T} f^q  +f \right], \label{eq:s} \\
  \varepsilon& = & g \int \frac{\dd^3p}{(2\pi)^3} m_{\rm T} f \label{eq:e} 
\end{eqnarray}

Here, $f$ is a momentum distribution function of the degrees of freedom (hadrons). In our case, it is defined through the Tsallis\,--\,Pareto dsitribution:
\begin{eqnarray}
  f(E,q,T_q,\mu) &= & \left[1+\frac{q-1}{T_q}(E-\mu)\right]^{-\frac{1}{q-1}}\label{eq:f} \ . 
\end{eqnarray}
Note that in the $q\to 1$ limit this formula reduces to the Boltzmann distribution, and provide the standard statistical thermodynamical definitions and function forms. In case of a mixed system, such as when unidentified hadron production is observed, a sum over the hadronic states has to be applied.

\begin{figure}[ht!]
  \centering    
  \includegraphics[width=\linewidth]{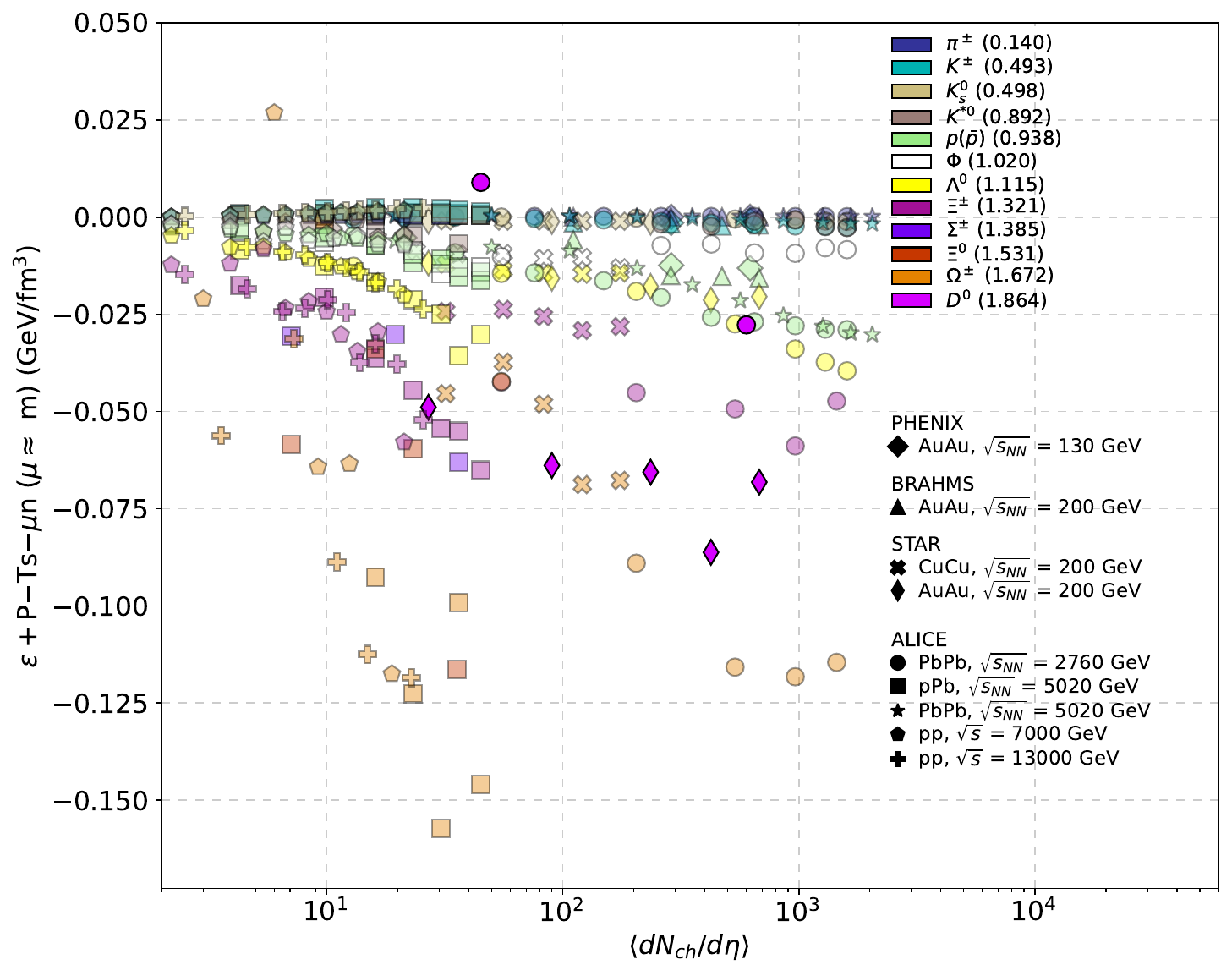}
  \caption{Thermodynamical consistency of the Tsallis--Pareto parameters. D-meson results (magenta) are compared to light-flavour hadron results from \cite{Biro:2020kve} (semi-transparent).}
  \label{fig:mult_ALL_c}
\end{figure}

Thermodynamical consistency is examined via the fulfilment of the first law of thermodynamics. We evaluate the validity of the non-extensive thermodynamical approach by checking the degree to which the Tsallis--Pareto distribution statisfies Eq.~(\ref{eq:consistency}). Figure~\ref{fig:mult_ALL_c} shows that the consistency is fulfilled for light-flavour mesons within 1\% precision. For heavier hadrons this can deviate slightly more, however, for D mesons it always stays below 8\%. 
Thus, for most of the hadrons that are measured with high precision, we observe thermodynamical consistency with the Tsallis description, and the results demonstrate scaling properties on light-flavor identified-hadron data from $\sqsn=130$ GeV to 13 TeV at mid-rapidity~\cite{Biro:2020kve}. The slight deviation in the consistency check observed in the heavy-flavor case may stem from the larger uncertainties of the D meson spectra, or mark the limits of the applicability of the non-extensive framework on the given system. Note that while extensive thermodynamics in itself is also thermodynamically consistent, Boltzmann fits present neither the scaling properties nor the thermodynamic consistency. Non-extensive thermodynamics is a generalization of the former which brings substantial improvement to the description of hadron spectra.

\newpage
\newcommand{\newblock}{}
\bibliography{CharmManuscript}


\end{document}